New experimental evidence for the role of long-range potential fluctuations in the mechanism of 1/$f$ noise in $a$-Si:H

J.P.R. Bakker[1], P.J.S. van Capel[1], B.V. Fine[2,3], and J.I. Dijkhuis[1]

[1]Debye Institute and [2]Spinoza Institute, Department of Physics and Astronomy, Utrecht University, P.O. Box 80000, 3508 TA Utrecht, the Netherlands.
[3]Max Planck Institute for the Physics of Complex Systems, Noethnitzer Str. 38, D-01187 Dresden, Germany (present address).

**Abstract**
We present measurements of 1/$f$ resistance noise in three different films of amorphous silicon ($a$-Si) in the presence of a transverse electric current. Two of these films have a *nin* sandwich structure – in one of them all three layers were hydrogenated; in the other one only the *n*-layers were hydrogenated, while the intrinsic layer was deuterated. The third film had *pip* structure with all three layers hydrogenated. The experimental spectra were found to be in a very good quantitative agreement with theoretical predictions, which were based on the mechanism involving long-range fluctuations of the Coulomb potential created by charged defects.

**I. Introduction**
We present new experimental tests of our recent theoretical proposal [1] explaining the origin of low frequency resistance noise in hydrogenated amorphous silicon ($a$-Si:H).
Usually, this kind of noise has a spectral shape close to 1/$f$, where $f$ is frequency. Our theory was fully developed for μm-thick films in the presence of a transverse electric current. It explained the resistance fluctuations by linking them to the thermal fluctuations of the occupation numbers of deep charged defects identified as dangling bonds. The noise mechanism involved the fluctuations of the long-range Coulomb potential created by those defects. The 1/$f$-like spectral shape was related to the distribution of potential barriers for the charge carrier escape from deep defects. This theoretical picture is amenable to a first principles calculation of the integrated noise intensity. It also yields the prefactor in front of the 1/$f$ frequency dependence. In the original work [1] we studied 1/$f$ noise in the sandwich structure consisting of electron doped – undoped – electron doped layers (*nin* structure) of $a$-Si:H. In that structure the resistance noise originated from the middle of the undoped layer, where electrons are the majority carriers.

The above mentioned theory implies two clear predictions. First, it should be directly generalizable to the case of *pip* structure, where holes are the majority carriers in the middle of the undoped layer. In Ref.[1], it was assumed that the distribution of the potential barriers for electron escape from deep defects is related to the medium range disorder of the local potential around the defects. This type of disorder should not be much different for the holes with respect to the valence band edge $E_V$ as compared with that of the electrons with respect to the conduction band edge $E_C$. Therefore, the natural expectation is that: (i) our theory should give correct 1/$f$ noise intensity for *pip* structures and (ii) if the samples are grown under identical conditions, the distribution of the potential barriers obtained from the studies of *pip* structures should be nearly the same (if counted from $E_V$). We test the above prediction on a *pip* structure, grown in the same reactor and under similar conditions as the *nin* structure studied in Ref. [1].

The second theoretical prediction is that, if the mechanism of charge fluctuations on deep defects is correctly identified in Ref. [1] with the escape and the capture of carriers, then the 1/*f* noise spectrum should not depend on the substitution of hydrogen with its heavier isotope deuterium - subject of the other test presented in this work

In the following we shall assume, that the reader is familiar with Ref. [1]. Therefore, will use the same set of variables and the same equations for the calculation of the theoretical spectra. To generalize the theory from Ref. [1] to the hole conduction noise in *pip* structures, the energy axis should be reversed, $E_C$ and $E_{C0}$ should be replaced with $E_V$ and $E_{V0}$, respectively, in all formulas, and effective electron mass should be replaced by effective hole mass. In this way, all variables describing the $D^+ \leftrightarrow D^0$ transition will describe the $D^0 \leftrightarrow D^-$ transition and *vice versa*.

**II. Devices**

|  | *nin B1* | *pip C* | *nin D1* |
|---|---|---|---|
| $d_i$ (μm) | 0.91 | 0.84 | 1.12 |
| $A$ (cm$^2$) | 0.55 | 0.57 | 0.57 |
| $n_D$ (cm$^{-3}$) | $6 \cdot 10^{15}$ | $6 \cdot 10^{15}$ | $4 \cdot 10^{15}$ |
| $E_0$-***m****(eV)* | -0.2 | 0.0 | -0.2 |
| ***b*** (eV/μm$^2$) | 1.59 | 1.54 | 1.24 |

Table 1. Devices parameters (defined in Ref. [1]).

The first device, *B1*, is the *nin* film of *a*-Si:H, the same as reported in Ref.[1]. The second device, *C*, is the *pip* structure of *a*-Si:H. The third device, *D1*, is the *nin* structure, where the undoped layer was deuterated, while the *n*-layers were hydrogenated. The amorphous silicon layers were all grown by Plasma Enhanced Chemical Vapor Deposition (PECVD) at an RF field of 13.56 MHz and a substrate temperature of 468 K. The intrinsic layer of thickness $d_i$ and area *A* was sandwiched between either ~ 50-nm-thick *n*-type layers or, in case of device *C*, 28-nm-thick microcrystalline *p*-type layers. Doped crystalline silicon wafers were used as a substrate, and the top contact consisted of Ti/Cu layers. Further details of deposition and devices can be found in Table 1 and Ref. [2], which uses the same device names.

The energy distribution of deep defects was modelled by two Gaussians, shifted with respect to each other by the correlation energy $U = 0.2$ eV. One of the Gaussians described the $D^+ \leftrightarrow D^0$ transition and the other one $D^- \leftrightarrow D^0$ transition. We applied the one-dimensional device simulation program AMPS [3] to fit to the measured current-voltage characteristics and thus obtained the concentration of deep defects $n_D$, the band bending coefficient ***b***, and the positions of the maximum of the first of the above Gaussians $E_0$. The best fit was obtained for the values of the defect parameters reported in Table 1. The half-width of the Gaussian distribution $\Delta E$ was fixed at 0.15 eV, the bandgap at 1.80 eV, the carrier mobility at 10 cm$^2$V$^{-1}$s$^{-1}$, the capture cross sections at 10$^{-15}$ cm$^2$, effective electron mass at 0.4 $m_e$, effective hole mass at $m_e$, and the density of states at the mobility edge at $4 \times 10^{21}$ eV$^{-1}$cm$^{-3}$. The values of $E_{C0}$ -**m** for the *nin* devices and $E_{V0}$ -**m** for the *pip* devices were obtained from the temperature dependence of the resistivity.

We measure the voltage noise spectra at frequencies in the range from 1 to 10$^4$ Hz and in the temperature range 340 K – 434 K. All calibrated spectra were taken at 50 mV, in the Ohmic regime, and are the result of averaging over 300 individual spectra. Further details of our experimental setup are described in Ref. [4].

**III. Experimental results**
(i) *nin a*-Si:H device. The experimental noise spectra taken from the *a*-Si:H *nin* device *B1* are presented in Fig. 1 together with the theoretical spectra computed on the basis of Eq. (27) from Ref. [1]. The integrated noise intensity is determined with no adjustable parameters from the numbers given in Table 1 and 2 and in Section II. The theoretically predicted noise spectra, which have a theoretical accuracy of a factor of two, are in good agreement with the measurements. The central position $E_{B0}$ of the Gaussian distribution of defects $P(E_B)$ and its width $\Delta E_B$ are the only adjustable parameters. They govern the curvature of the noise spectra. The effective distribution of barriers $P(E_B) r_s^3(E_B)$ shown in Fig.1(d) is asymmetric. It peaks at the barrier energy $E_{Bdist}$, which is shifted with respect to the peak of the Gaussian distribution of $P(E_B)$ by approximately 0.1 eV. For this device (and for the two others), the peak of the effective distribution $E_{Bdist}$ coincides with the experimental activation energy $E_{Bexp}$ of the characteristic noise frequency (defined in Ref. [4]), as it should.

(ii) *pip a*-Si:H device. The noise mechanism for holes is tested using the *pip a*-Si:H device. The qualitative behavior of the noise spectra in Fig. 2 turns out to be analogous to that of *nin* devices. The theoretical curves were calculated using Eq. (27) of Ref. [1] but now for the hole transitions to and from charged deep defects. Because of the larger conductivity activation energy $E_\mathbf{s}$, the input value for $E_{B0}$ - **m** in *pip* devices is 0.15 eV larger than the corresponding value for *nin* devices (see Table 2 and inset of Fig. 2). However, when counted from the valence mobility edge, the value of $E_{B0}$-$E_{V0}$ is very close to $E_{B0}$ - $E_{C0}$ for the *nin* device *B1*. The values of $\Delta E_B$ for the two devices are also in reasonable agreement.

|  | *nin B1* | *pip C* | *nin D1* |
|---|---|---|---|
| $E_\mathbf{s}$ (eV) | 0.63 | 0.82 | 0.67 |
| $E_{Bexp}$ (eV) | 0.81 | 0.94 | 0.88 |
| $E_{Bdist}$ - **m**(eV) | 0.81 | 0.98 | 0.86 |
| $E_{B0}$ - **m**(eV) | 0.90 | 1.05 | 0.99 |
| $E_{B0}$ - $E_{C0}$ (eV) | 0.27 | 0.23 | 0.32 |
| $\Delta E_B$ (eV) | 0.09 | 0.07 | 0.11 |

Table 2. Measured and fitted energy barriers and width.

(iii) *nin a*-Si:D device. Typical spectra for device *D1* (passivated with deuterium) are shown in Fig. 3. There is hardly any difference between noise spectra from devices with hydrogen or deuterium in the temperature range from 295 K to 350 K. Electronically, the devices are identical, but have a 23 % difference in thickness (see Table 1). In the theory, this results in a slightly higher noise intensity for the deuterium device predominantly in the low-frequency region in the temperature range from 350 K to 434 K. Given the agreement between experiment and theory, we conclude that the main reason for the small difference of the curvature between the two devices lies in the difference in thickness, which induces an increase of the screening radius in the deuterium device. The barrier distribution for this device is characterized by the values of $E_{B0}$-$E_{C0}$ and $\Delta E_B$, which are close to the values of the same parameters obtained for the hydrogenated device.

**IV. Conclusion**

We have measured the 1/$f$ noise spectra in *nin* and *pip* *a*-Si:H devices and in an *nin* *a*-Si:D device. In all three cases, we have found a very good quantitative agreement between the experiments and the theory proposed in Ref. [1]. The theory is valid both in the regime, where electrons are the majority carriers (*nin* devices), and in the regime, where holes are the majority carriers (*pip* device). The comparison of the results gives further support for the existence of a universal distribution of emission barriers in *a*-Si:H/D for the carrier escape from deep defects.

The maximum of that distribution is located approximately 0.27 eV above the conduction band or valence band mobility edges for electrons and holes, respectively. The width of this distribution is about 0.09 eV. Comparing the results for the deuterated and hydrogenated devices, we observed no significant difference, which was also consistent with the theory.

Caption Figure 1.
Noise spectra measured in the Ohmic regime of the *I-V* curve in a). Erratic lines indicate experimental data and smooth lines predictions of Eq. (27) of Ref. [1]. The spectra at 340 K, 362 K and 402 K are multiplied, respectively, by $30^3$, $30^2$ and 30 to make them distinguishable. In (b), the Gaussian distribution of emission barriers, with the chemical potential μ set to zero. The screening radius is displayed in (c), and the effective distribution of barriers in (d).

Caption Figure 2.
Noise spectrum of *pip* device *C* measured at 434 K together with the theoretical predictions. Inset: effective distributions of barriers for *nin* and *pip* devices.

Caption Figure 3.
Measured (symbols) and theoretical (lines) noise spectra of deuterated *nin* device *D1* measured at 402 K and at 340 K, the latter being multiplied by a factor of 30 for clarity.

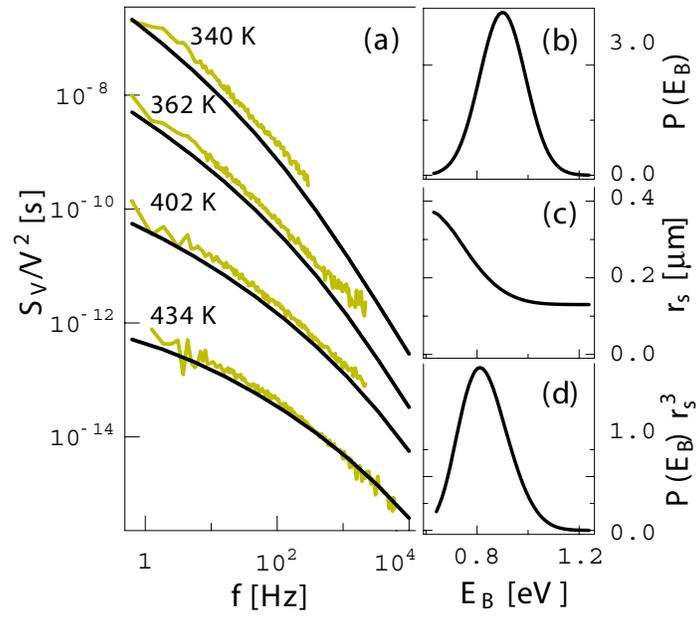

Fig. 1

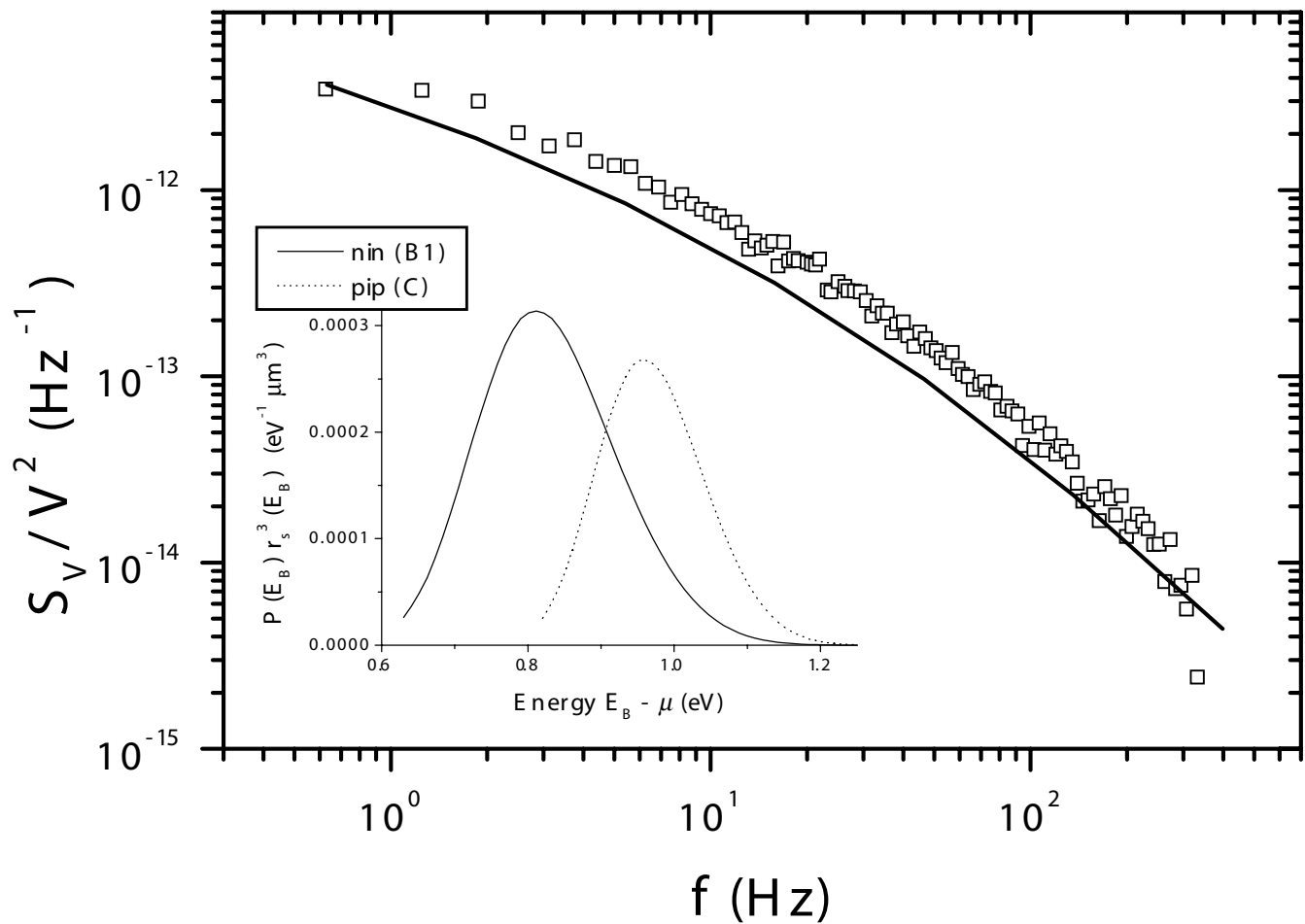

Fig. 2

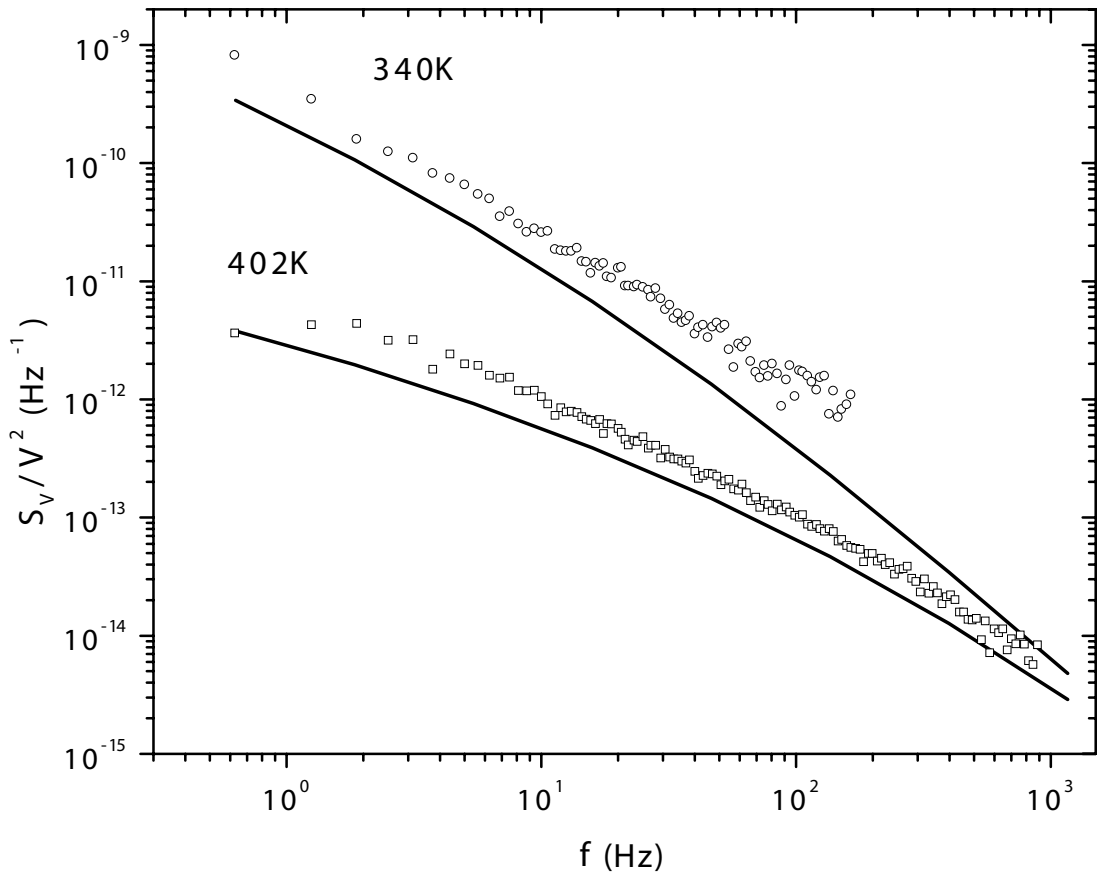

Fig. 3